\documentstyle[aps,prl,epsfig,twocolumn]{revtex}

\newcommand{\be}{\begin{equation}}
\newcommand{\ee}{\end{equation}}
\newcommand{\bea}{\begin{eqnarray}}
\newcommand{\eea}{\end{eqnarray}}
\newcommand{\bdm}{\begin{displaymath}}
\newcommand{\edm}{\end{displaymath}}
\newcommand{\bdma}{\begin{eqnarray*}}
\newcommand{\edma}{\end{eqnarray*}}
\newcommand{\ba}{\begin{eqnarray*}}
\newcommand{\ea}{\end{eqnarray*}}
\newcommand{\bi}{\begin{itemize}}
\newcommand{\ei}{\end{itemize}}
\newcommand{\benum}{\begin{enumerate}}
\newcommand{\eenum}{\end{enumerate}}

\input{colors.sty}  

\newcommand{\refkl}[1]{(\ref{#1})}

\newcommand{\erw}[1]{\mbox{$\langle #1 \rangle$}} 

\newcommand{\sub}[1]{_{\rm #1}}

\begin{document}

\tighten
\twocolumn[\hsize\textwidth\columnwidth\hsize\csname @twocolumnfalse\endcsname
{\protect

\title{Memory effects in microscopic traffic models and wide scattering in flow-density data}
\author{Martin Treiber and Dirk Helbing}
\address{Institute for Economics and Traffic, Dresden University of Technology,
Andreas-Schubert-Str. 23, 01062 Dresden, Germany
}
\date{\today}
\maketitle
\begin{abstract}
By means of microscopic simulations
we show that non-instantaneous adaptation of the driving behaviour to the
traffic situation together with the 
conventional measurement method of flow-density data 
can explain the observed inverse-$\lambda$ shape and the
wide scattering of flow-density data in ``synchronized'' congested
traffic. We model a memory effect in the response of drivers to the traffic situation 
for a wide class of car-following models
by introducing a new dynamical variable describing
the adaptation of drivers to the surrounding traffic
situation during the past few minutes (``subjective level of
service'') and couple  this internal state to
parameters of the underlying model that are related to the driving style.
For illustration, we use the intelligent-driver model (IDM) 
as underlying model, characterize the level of service solely 
by the velocity and couple the internal variable 
to the IDM parameter ``netto time gap'', 
modelling an increase of the time gap in congested traffic
(``frustration effect''),
that is supported by single-vehicle data.
We simulate open systems with a bottleneck and obtain
flow-density data by implementing
``virtual detectors''. Both the shape,
relative size and apparent ``stochasticity'' of the region 
of the scattered data points 
agree nearly quantitatively with empirical data.
Wide scattering is even observed for identical vehicles, although
the proposed model is a time-continuous,
deterministic, single-lane 
car-following model with a unique fundamental diagram.
\end{abstract}


} ]   

\section{Introduction}

The nature of ``synchronized'' traffic flow
is one of the most controversial subjects in traffic theory
\cite{Helb-Opus,TGF01}.
It is a form of congested traffic with nonzero flows typically
found upstream of inhomogeneities 
(e.g. freeway bottlenecks), characterized by an erratic motion of
time-dependent flow-density data in a two-dimensional area 
(and a synchronization
of the time-dependent average vehicle
velocities among neighboring lanes) \cite{KeRe96,Kerner-sync,Ker02_PRE}. 

The wide scattering of the data points for congested traffic seems to 
exclude explanations in terms of traffic models 
assuming a fundamental (steady-state) relation 
$Q_{\rm e}(\rho)$ between the flow $Q$ and the density $\rho$.
In response, models with non-unique flow-density relations (or
velocity-distance relations) have been proposed both on a
macroscopic level \cite{Nelson-Sync},
as car-following models
\cite{Wiedemann,Namazi-02,Kerner-Mic,Helb-SyncStock},
and as cellular automata \cite{Kerner-CA}.
The empirical data scattering has also triggered a flood of publications in 
physics journals with various other suggestions ranging from
shock waves propagating forward or backward
\cite{KeRe96}, effects of lane changing,
changes in the behavior of ``frustrated'' drivers 
\cite{Brilon-traff95,Kra98a,Treiber-Pinch}, anticipation effects
\cite{Lenz-Wagner,Kno01}, or a trapping of vehicles \cite{Lub02}.
Another obvious explanation of the scattering lies in the heterogeneity of
vehicles (such as cars and trucks) and driving styles (such as defensive or
aggressive) on any real road \cite{Banks-scatter}.
In fact, statistical analyses of single-vehicle data show a particularly wide
scattering of the time gaps between successive vehicles in congested traffic
\cite{Tilch-TGF99,Neubert-TGF99,Katsu03-condmat}.
Furthermore, the average {\em netto} time gaps increase in congested traffic
suggesting that ``frustration effects'' are real \cite{Katsu03-condmat}.
Macroscopic simulations taking into account
observed variations in the truck percentage \cite{GKT-scatter} or direct
microsimulations 
with two types of vehicles \cite{Treiber-TGF99} could explain a great deal
of the observed scattering, but the 2D-regions
remained somewhat smaller than in the observed data.

Another factor  possibly contributing to the wide scattering 
are traffic instabilities resulting in so-called ``oscillating
congested traffic'' which is the most common form of congested traffic
\cite{IDM}. If the sampling time interval for data aggregation is 
not commensurable with the frequency of the oscillations or if the
oscillations are nonperiodic, then the
data points will display artificial ``erratic scattering''.
In this case, the origin of the scattering is the method of  
data aggregation in combination with the conventional
interpretation of flow-density data \cite{Banks-scatter}.

In this paper we show by means of simulations that
the adaptation of drivers to the surrounding 
traffic on time scales of a few minutes (``memory
effect'') in conjunction
with traffic instabilities can quantitatively explain the observed scattering.
Our model is based on the observation that,
after being stuck for some time 
in congested traffic, most drivers 
adapt their driving style, e.g., by 
increasing their preferred {\it netto} (bumper-to-bumper)
time gap $T$ to the preceding vehicle
\cite{Tilch-TGF99,Neubert-TGF99,Katsu03-condmat,TGF01}. 
Apart from congestion, other aspects of the traffic environment
such as driving in the dark or in tunnels affect the driving behaviour as well
\cite{Edie-tunnel}, but will not considered in this paper.

Models based on memory effects have been successfully applied in 
several fields of statistical and interdisciplinary physics
such as liquid crystals and polymers
\cite{DeGennesPoly} or  in
reaction-diffusion systems \cite{Fedotov02}.
Nonlinear Markow equations with memory Kernel have been applied
to multi-agent systems and financial markets and provide a
statistical mechanism for the observed clustered volatility \cite{Schulz01}.

In the next section, we formulate the adaptation of the driving style to the
surrounding traffic and incorporate the memory effects into
the intelligent-driver model (IDM) \cite{IDM} resulting in the IDMM
(``intelligent-driver model with memory'').
Like the IDM, the IDMM is a deterministic 
time-continuous car-following model with a unique steady-state
flow-density relation. 

In Section \ref{sec_sim}, we presents related simulations and compare the
measurements of ``virtual detectors'' with empirically measured traffic data. 
We will find a semi-quantitative agreement.
While the scattering is even observed in the 
original IDM, the memory effect is
necessary to obtain the inverse-$\lambda$ shape and the correct relative size of the 
twodimensional region of data scattering.

In the concluding Section \ref{sec_diss} we discuss the remarkable
fact that erratic scattering  can be obtained from deterministic
single-lane models
with a unique fundamental diagram in
simulations without any element of stochasticity or heterogeneity. 

\begin{figure}
\begin{center}
\includegraphics[width=85mm]{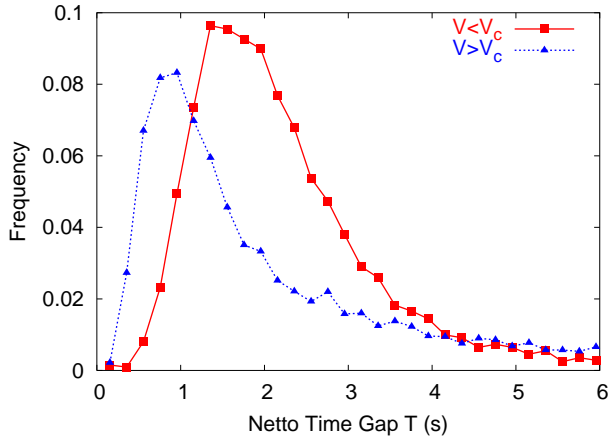} 
\end{center}
\caption[]{Distribution of netto time headways on the left lane of the
Dutch A9 from Haarlem to Amsterdam for
free traffic ($v > v_c=60$ km/h) and congested traffic 
($v<v_c$).
}
\label{fig_Tdistr}
\end{figure}

\section{Model equations}

We will formulate the memory effect in human driver behavior in a way that directly
connects to existing car-following models.
Any model can be used, the parameters of which can be interpreted in terms of the
driving behaviour. The model should allow to define some desired velocity
$v_0$ and to influence the minimum netto time gap $T$ 
by varying one or more
of its model parameters. The outcome, of course, will depend on the
details of the model used.
In this paper, we apply the intelligent-driver model (IDM) \cite{IDM}
as underlying model, where $v_0$ and $T$ are 
model parameters themselves.

\subsection{The intelligent-driver model}

In the IDM, the acceleration of each vehicle $\alpha$ is assumed to be a continuous function 
of the velocity $v_{\alpha}$, the netto distance gap $s_{\alpha}$,
and the velocity difference (approaching rate)
$\Delta v_{\alpha}$ to the leading vehicle:

\begin{equation}
\label{IDMv}
\dot{v}_{\alpha} = a
         \left[ 1 -\left( \frac{v_{\alpha}}{v_0} 
                  \right)^4 
                  -\left( \frac{s^*(v_{\alpha},\Delta v_{\alpha})}
                                {s_{\alpha}} \right)^2
         \right].
\end{equation}
This expression is {an interpolation of the tendency to accelerate 
with 
$a_f(v) := a[1-(v/v_0)^4]$ 
on a free road and the tendency to brake with deceleration
$-b_{\rm int}(s, v, \Delta v)
:= -a(s^*/s)^2$, when vehicle $\alpha$ comes too
close to the vehicle in front.} The deceleration term
depends on the ratio between the effective ``desired
minimum gap'' $s^*$ and the actual gap $s_\alpha$, where the desired gap
\begin{equation}
\label{sstar}
s^*(v, \Delta v) 
    = s_0 
    + v T
    + \frac{v \Delta v }  {2\sqrt{a b}}
\end{equation}
is dynamically varying with the velocity.
The first term $s_0$ on the right-hand side denotes the small 
minimum distance kept in standing traffic.
The second term corresponds to following the preceding vehicle
with a constant ``safety'' netto time gap $T$.
The third term is only active in non-stationary traffic 
and implements an accident-free ``intelligent'' driving behaviour
including a braking strategy that, in nearly all situations,
limits braking decelerations to the ``comfortable
deceleration'' $b$. 


\subsection{Adaptation of the driving style and memory effect}

We assume that all adaptations of the driving style 
are controlled by a single internal dynamical
variable $\lambda_{\alpha}(t)$ (``subjective level of service''), 
which can take on values between
0 (in standing traffic) and 1 (on a free road), and that it relaxes to
the instantaneous level of service $\lambda_0(v)$ 
with a relaxation time $\tau$ according to
\be
\label{lambda}
\frac{d\lambda_{\alpha}}{d t} 
= \frac{\lambda_{\alpha}-\lambda_0(v_{\alpha})}{\tau}.
\ee
This means, for each driver the subjective level of service is given
by the exponential moving average (EMA) of the instantaneous level of service
experienced in the past: 
\be
\label{lambdaexp}
\lambda_{\alpha}(t)=\erw{\lambda_{0\alpha}}\sub{EMA}
= \int_{0}^t \lambda_0(v_{\alpha}(t')) e^{-(t-t')/\tau} \ dt' .
\ee

We have assumed the instantaneous level of service $\lambda_0(v)$
to be a function of the actual velocity $v(t)$. Obviously,
$\lambda_0(v)$ is a monotonuously increasing 
function with $\lambda_0(0)=0$ and $\lambda_0(v_0)=1$.
In this paper, we specify the most simple ``level-of-service function''
satisfying these conditions:
\be
\label{lambda0}
\lambda_0(v) =\frac{v}{v_0}.
\ee
Notice that this equation reflects the level of service or efficiency of movement from the driver's point
of view, with $\lambda_0=1$ meaning zero hindrance and $\lambda_0=0$
meaning maximum hindrance. If one models heterogeneous traffic 
where different drivers have different
desired velocities there is no ``objective'' level of service, only an average one.

Having defined how the traffic environment influences the 
degree of adaptation $\lambda_{\alpha}$ of each driver, we now 
specify how this internal variable
influences the driving behaviour. A behavioural variable that is both
measurable 
and strongly influencing the traffic dynamics is the
netto time gap $T$.
Figure \ref{fig_Tdistr} shows that, in congested traffic,
the whole distribution of time gaps is
shifted to the right compared to data of free traffic \cite{Katsu03-condmat}. 
We model this by varying the corresponding 
IDM parameter in the range between $T_0$ (free traffic) and
$T\sub{jam}=\beta_T T_0$ (traffic jam) according to
\be
\label{Tdyn}
T(\lambda)=T_0[\beta_T+\lambda (1-\beta_T)] \, .
\ee
Herein, the {\it adaptation factor} $\beta_T$ is a model parameter
(cf. Table \ref{tab_param}). 
Notice that probably other parameters of the driving style are influenced as well,
such as the acceleration, the
comfortable deceleration, or the
desired velocity. This could be implemented by
analogous equations for $a$, $b$, and $v_0$,
respectively.
For simplicity (and in order to have an empirically testable model),
we will only consider the influence on $T$.


\newcommand{\entry}[2]{\parbox{55mm}{#1} &
                      \parbox{25mm}{#2} }

\begin{table}

\begin{tabular}{l|l}
\entry{Parameter} {Typical value}  \\[3mm] \hline 
\entry{}{}{} \\[-2mm]
\entry{Desired velocity $v_0$}
     {120 km/h}
     \\[0mm]
\entry{Netto time gap $T_0$}
     {0.85 s}
     \\[0mm]
\entry{Maximum acceleration $a$}
     {0.8 m/s$^2$}
     \\[0mm]
\entry{Comfortable deceleration $b$}
     {1.8 m/s$^2$}
     \\[0mm]
\entry{Minimum distance $s_0$}
     {1.6 m}
     \\[0mm]
\entry{Vehicle length $l = 1/\rho_{\rm max}$}
     {6 m}
     \\[3mm]
\entry{Adaptation factor $\beta_T = T\sub{jam}/T_0$}
     {1.8}
     \\[0mm]
\entry{Adaptation time $\tau$}
     {600 s}
     \\[0mm]
\end{tabular}
\vspace*{5mm}
\caption[]{\label{tab_param} Model parameters of the IDMM
with the values used throughout this paper. All eight model parameters
have a clear vehicle- or driver-related meaning and can be determined
from empirical traffic data. Note that other models recently proposed
to describe similar empirical observations consist of up to
10 equations with about
20 parameters \protect\cite{Kerner-Mic}.
}
\end{table}

In summary, the IDMM
is described by the IDM equations \refkl{IDMv} and \refkl{sstar},
by Eq. \refkl{Tdyn} describing how the subjective
level of service $\lambda$ influences the time gap, and by the 
dynamical equation for the internal state itself, which can be written as
\be
\label{lambdaSpec}
\frac{d\lambda_{\alpha}}{d t} = \frac{\lambda_{\alpha}-v_{\alpha}/v_0}{\tau}.
\ee
All IDMM parameters
are intuitive and can be determined from traffic data.
In the special case $\beta_T=1$, the IDMM reverts to the original IDM.
The special case $\tau=0$ corresponds to a slightly modified  IDM, 
where the parameter $T$ in Eq. \refkl{sstar} is replaced by
$T(v)=T_0[\beta_T+\frac{v}{v_0} (1-\beta_T)]$.
Table  \ref{tab_param} gives the values that we will used throughout the rest
of this paper unless stated otherwise.

\begin{figure}
\begin{center}
\includegraphics[width=80mm]{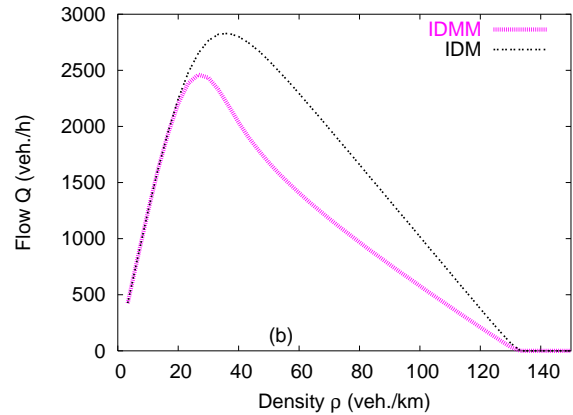} 
\end{center}
\caption[]{
Comparison of the theoretical fundamental diagrams of the 
IDM and IDMM.
}
\label{fig_fundIDMM}
\end{figure}

Notice that the IDMM belongs to the class of models with a unique
stationarity relation. Its steady-state following distance
as a function of the velocity is given by
\be
\label{se}
s_e(v)
  = \frac{s_0+v T_0 \left(\beta_T + (1-\beta_T) \ \frac{v}{v_0} \right)}
{\sqrt{1-\left(\frac{v}{v_0}\right)^4}} \, .
\ee
Figure \ref{fig_fundIDMM} shows the resulting fundamental diagram for
identical vehicle types for the IDMM in comparison with that of the IDM.

\section {Simulations}
\label{sec_sim}
We have simulated a 20 km long road section with a bottleneck
and open boundaries, 
assuming identical vehicles of length $l=6$ m
whose drivers behave according to the IDMM with
the parameters given in Table \ref{tab_param}.
The simulations have been started with very light traffic 
correspondig to a homogeneous density of 2 vehicles/km 
and an initial velocity of 100 km/h. 
During the simulated time interval of
3 hours, we have simulated idealized rush-hour conditions by
increasing the inflow at the upstream boundary linearly from
200 veh./h at $t=0$ to 2400 veh./h at $t=25$ min. Afterwards,  the flow
has been decreased
linearly to  100 veh./h at $t=180$ min.
All vehicles have been initialized with $\lambda=1$, i.e., with a ``memory'' of
free traffic.
As in macroscopic traffic simulations of open systems
\cite{numerics},
the velocity of the inflowing
traffic turned out to be irrelevant, since it quickly approached the value
corresponding to the ``free'' branch of the velocity-flow
relation $v_e(Q)$ with a flow $Q(t)$ equal to that imposed at the boundary.

We have implemented a flow-conserving bottleneck 
by locally increasing the IDM parameter $T_0$,
\be
\label{T0}
T_0=T_0(x)=\left\{ \begin{array}{ll}
  1.20 \mbox{ s} & 17 \ \mbox{km} \le x<18 \ \mbox{km}, \\
0.85 \mbox{ s} & \mbox{otherwise.}
\end{array} \right.
\ee
This corresponds to lowering
the road capacity and can represent any bottleneck which is not an on-
or off-ramp \cite{Helb-micmac}.
Notice that, at the bottleneck, this means that the actual 
time headway $T$, as specified by Eq. \refkl{Tdyn} with \refkl{T0},
depends both directly on $x$ and on the subjective level 
of service of the driver resulting in another indirect dependence on
$x$.

The simulation has been performed using an explicit integration scheme. 
Output is produced by implementing ``virtual detectors'' at $x=9$ km
and $x=12$ km, with data aggregation periods of $T\sub{aggr}=60$ s.
In each aggregation interval $i$, the traffic flow 
\be 
Q_i=n_i/T\sub{aggr}
\ee
is determined by counting the number $n_i$ of crossing vehicles, 
the average velocity $V_i$ is calculated by the 
arithmetic average 
\be
V_i=\frac{1}{n_i} \sum_{\alpha=1}^{n_i} v_{\alpha} ,
\ee
and the density by 
\be
\label{rhoemp}
\rho_i=Q_i/V_i,
\ee
as in many practical cases.
\begin{figure}
\begin{center}
\includegraphics[width=70mm]{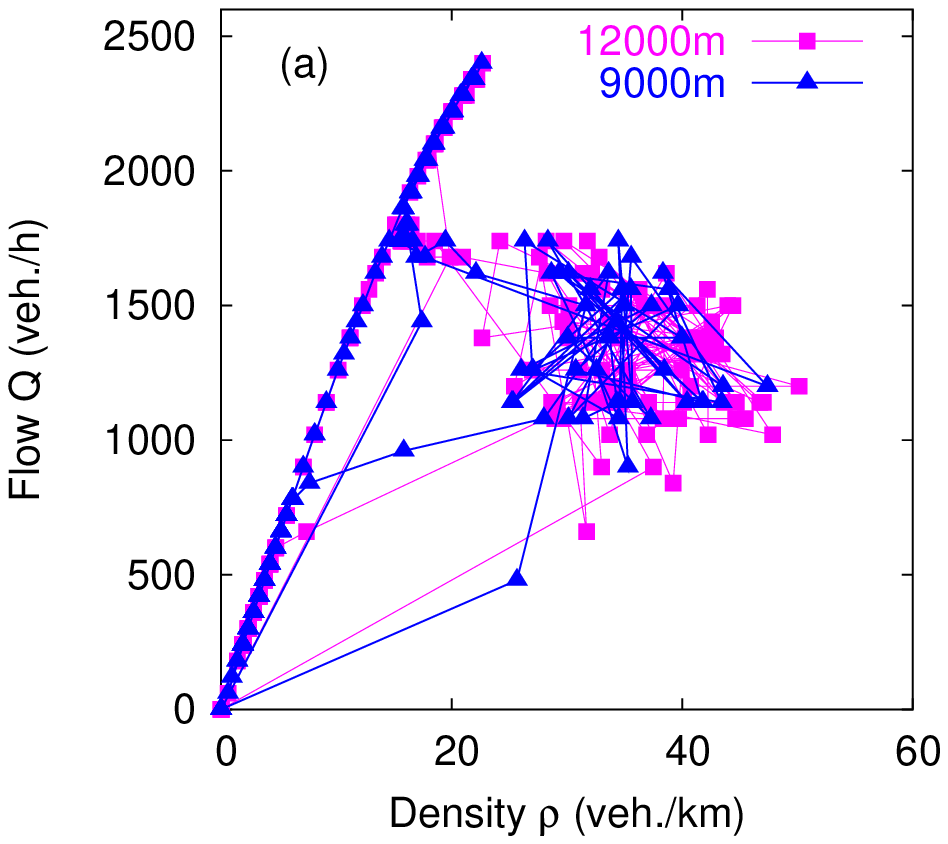} 
\includegraphics[width=70mm]{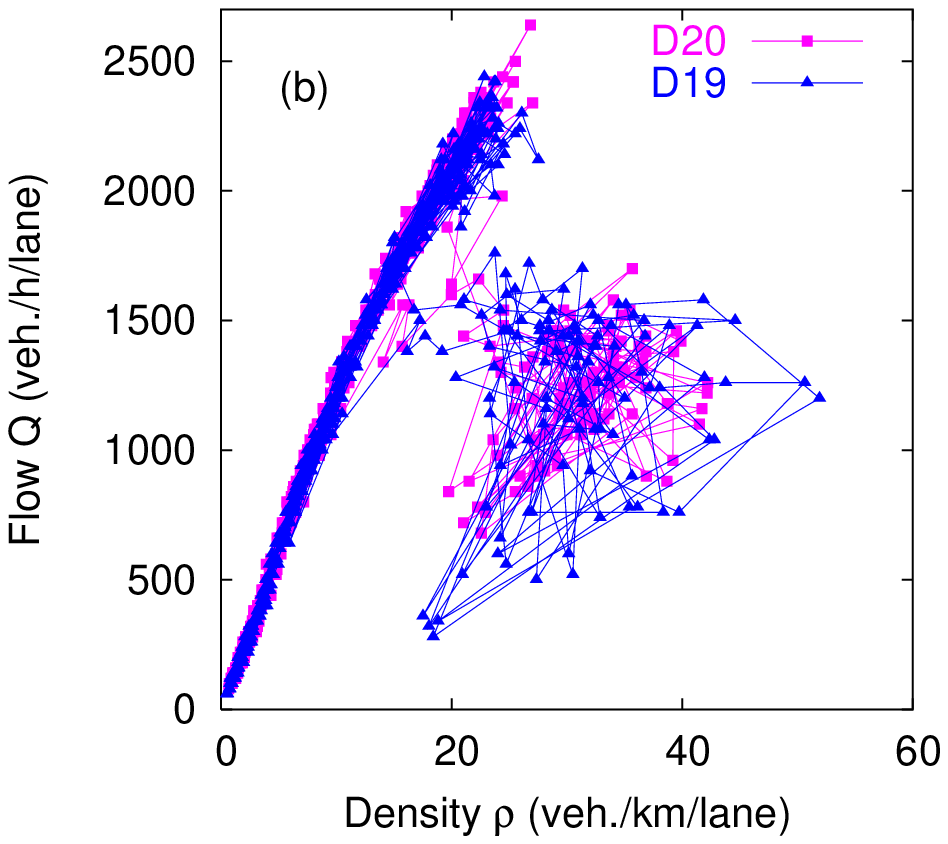} 
\includegraphics[width=65mm]{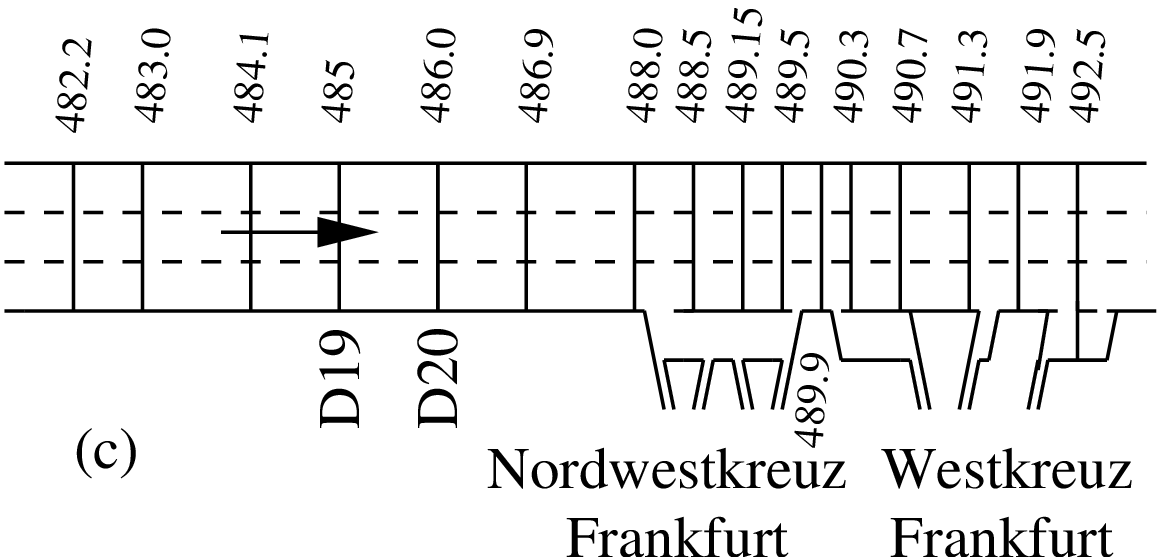} 
\end{center}
\caption[]{(a) Simulated flow-density data of two virtual detectors
compared with (b) flow-density data from the German freeway
A9-South near Frankfurt on July 31, 2001. (c) Location of the 
detectors for the empirical data. Here, the 
intersection ``Nordwestkreuz'' serves as bottleneck. 
}
\label{fig_fund}
\end{figure}
%
Figure \ref{fig_fund} displays the resulting 
flow-density data of the two virtual detectors, compared with empirical
data from real traffic. Both diagrams shows 
(i) the characteristic wide and erratic scattering of
the data points which is a signature of ``synchronized traffic'',
(ii) the characteristic inverse-$\lambda$ shape with 
a maximum $Q\sub{max}$ 
of traffic flow in free traffic (immediately prior to
the breakdown), which is distinctively higher than the typical flows in
the congested traffic $Q\sub{CT}$ after breakdown.


\subsection{Interpretation of macroscopic traffic data}
The question arises
what causes the obvious stochasticity in the data of the 
virtual detectors although
everything in the simulation is deterministic (including the
upstream boundary condition and the implementation of the bottleneck),
although all drivers and vehicles are treated identically,
and not even lane changes may serve as possible source for fluctuations.
The only possible source of the fluctuations are traffic instabilities
which we will analyse in the following.
\begin{figure}
\begin{center}
\includegraphics[width=80mm]{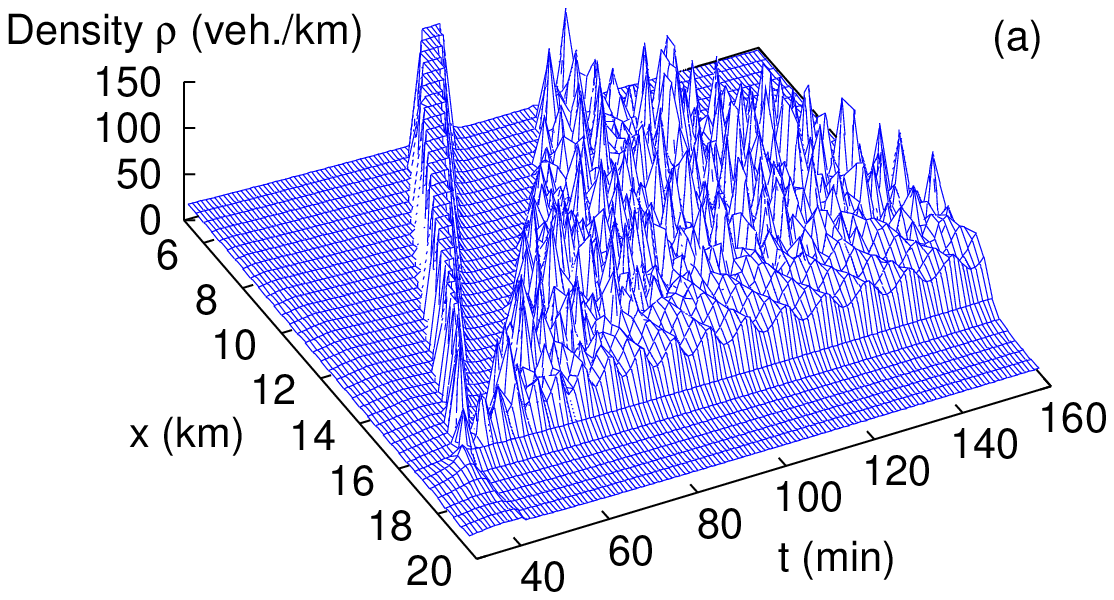} 
\includegraphics[width=80mm]{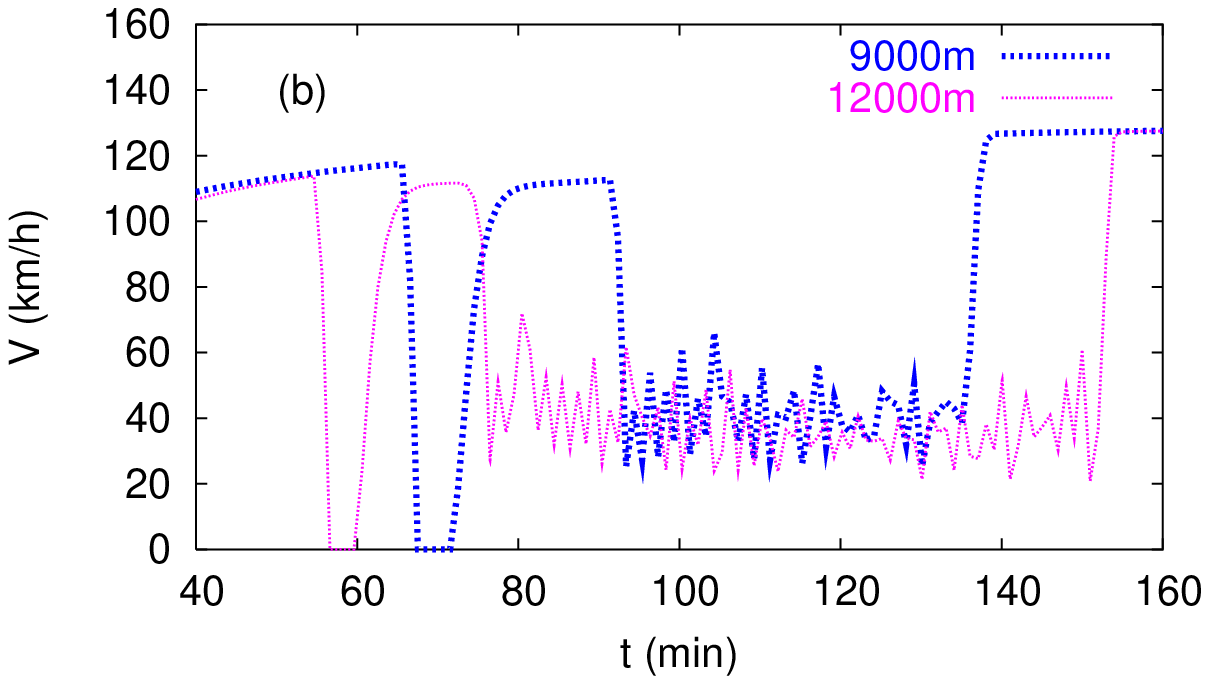} 
\end{center}
\caption[]{
(a) Spatio-temporal dynamics of the traffic density 
and (b) time series of the velocity at the two virtual detectors
of the simulation shown in Fig. \protect\ref{fig_fund}.

}
\label{fig_rho3d}
\end{figure}
%
Figure \ref{fig_rho3d}(a) shows that at about $t=40$ min,
a traffic breakdown
occurs near the bottleneck  at $x=12$ km, triggering one isolated
wide jam with zero flow and a region of congested traffic with nonzero flow
upstream of the bottleneck. 
In the congested region, initially small oscillations behind the bottleneck
(oscillating congested traffic, OCT)
increase their amplitude while travelling further
upstream, and finally some isolated stop-and-go waves 
(called ``wide moving jams'' in \cite{KernerPinch})
are emitted from the congested region travelling further upstream
(triggered stop-and-go traffic \cite{Phase}).
The time series of the virtual detectors
(see Fig. \ref{fig_rho3d}(b)) show that the wide jam also crosses the
detectors. 
Since the jam and the oscillations of the OCT state 
 are  the only possible sources of
fluctuations, the 
scattering in the data of the virtual detectors obviously can
be traced back to longitudinal instabilities
in connection with the interpretation of the macroscopic data. 

Based on theoretical investigations\cite{KeRe96}, one might expect
a ``jam line'' in the flow-density diagram of Fig. 
\ref{fig_fund} stemming from the wide jam crossing the virtual detectors.
However, the jam line is missing. 
Moreover, the highest ``measured'' density is only about
50 veh./km, although the model parameters (Table \ref{tab_param})
imply a jam density of at least $\rho\sub{jam}=1/(l+s_0)=130$ veh./km.

\begin{figure}
\begin{center}
\includegraphics[width=70mm]{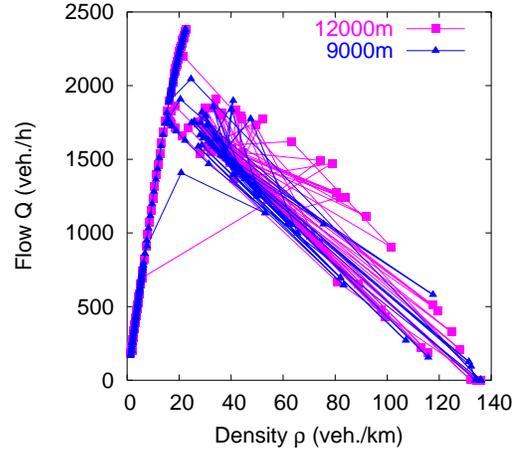} 
\end{center}
\caption[]{
Flow-density plot as in Fig.
\protect\ref{fig_fund}, but using no data aggregation and the real 
density \protect\refkl{rhoTrue}.
}
\label{fig_trueDensity}
\end{figure}
%
To check, whether this is an  artefact of the
data interpretation, in Fig.~\ref{fig_trueDensity} we have plotted 
the flow-density diagram of the same simulation
at the same locations, but this time 
using the true spatial density 
\be
\label{rhoTrue}
\rho(x,t)=\frac{1}{x_{\alpha-1}(t)-x_{\alpha}(t)}, \ \ 
x_{\alpha} \le x < x_{\alpha-1},
\ee
which one would obtain by ``snapshots'' 
at fixed times rather than the macroscopic density \refkl{rhoemp}.
In the data for $x=9$ km, one clearly sees the signatures
of the fully developed jam in form of a straight ``jam line'' $J$,
connecting the points $(\rho\sub{jam}, Q\sub{jam})=$ (130/km, 0) and
$(\rho\sub{out}, Q\sub{out})=$ (18 veh./km, 1750 veh./h).
Notice that, in accordance with observations \cite{KernerPinch},
the outflow from isolated jams is distinctively lower than
$Q\sub{max}$.
Moreover, the data at both locations show more than one instance
of zero or nearly zero flow at densities near the
maximum density, which can be seen neither in the flow-density data nor in
the time series of the velocity of the virtual detectors, cf. Fig.
\ref{fig_rho3d}. 
The reason is that the virtual detectors display 
finite velocities and flows whenever at least one car crosses the
detector during the sampling time interval. 
Thus, periods of standing traffic of
up to the double sampling time, i.e., up to 2 minutes,  may
not be observed in the detector output.

If we assume that the simulation captures some 
essential aspects of real traffic 
we conclude that 
(i) a jam line probably exists in real traffic but cannot be found in
flow-density data of stationary detectors,
(ii) when looking only at aggregated traffic data,  one might get 
a wrong picture of the actual traffic dynamics.

\subsection{Analysis of the adaptation effect}

We now proceed to investigate the effects of the new IDMM parameters.
We have simulated the
same system of equations with various values of the adaptation factor 
$\beta_T$ and the adaptation time $\tau$.
It turned out that both the values of
$Q\sub{out}$ and $Q\sub{CT}$ decrease with 
$\beta_T$, while $Q\sub{max}$ essentially 
remains unchanged. 
This is plausible, since $(\beta_T-1)$ describes the strength of the
``frustration effect'' after driving in congested traffic for some time, while
the value of $Q\sub{max}$ is related to free traffic where frustration effects
play a minor role. Furthermore, $Q\sub{CT}$ decreases with $\tau$.
Since the time spent in congestion behind bottlenecks
is typically of the order of the adaptation time or longer, 
drivers are adapted to congested traffic when they get closer to the downstream front of
the congestion area near the bottleneck. Consequently, it 
takes some time to revert to the more
``aggressive'' driving style in free traffic, explaining the decrease of
$Q\sub{out}$ with $\tau$. 

\begin{figure}
\begin{center}
\includegraphics[width=70mm]{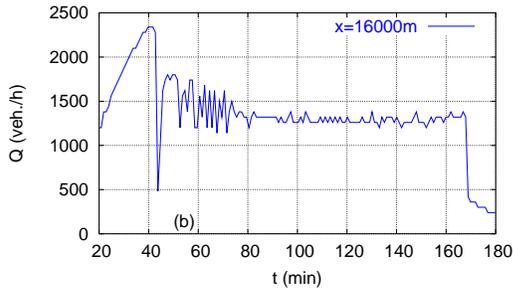} 
\end{center}
\caption[]{
Time series of traffic flow for the simulation of 
Fig. \protect\ref{fig_rho3d}, measured by a
virtual detector at $x=11000$ m near the bottleneck. 
Notice that the flow $Q\sub{out}\approx$ 1750 veh./h 
in the interval 40 min $\le t \le 50$ min
is related to the outflow from the isolated jam and not the bottleneck. 
}
\label{fig_q}
\end{figure}
%
One might argue that the drivers should adapt instantaneously to the traffic
situation. 
There is empirical evidence, however,
that the characteristic time scale of the
adaptation is not negligible \cite{TGF01}: In data of congested traffic
measured near the bottleneck causing the breakdown, one often observes that,
after the initial drop of the traffic flow, 
the flow decreases further during the first 10 or 20  minutes after
the breakdown, cf. e.g., Fig. 12 in \cite{IDM}.
However, according to traffic theory, 
the outflow from jams and thus the measured flow near the
bottleneck are constant for fixed driving styles.
Assuming that, 
after the breakdown, the length of the congested area behind the bottleneck
and thus the waiting time of each driver
increases, gradual adaptations naturally explain the observations.
Figure \ref{fig_q} shows this effect in the simulated measurement of a virtual detector
near the downstream congestion front. The average flow is highest
in the beginning (at $t=60$ min) and at the end (at $t=160$ min) 
of the congestion. However, it eventually decreases from $Q\sub{CT}\approx$
1500 veh./h to 
$Q\sub{CT}\approx$ 1300veh./h at $t=120$ min, where the length of the congestion
reaches its maximum value of about 10 km.

\subsection{Comparison with the IDM}

\begin{figure}
\begin{center}
\includegraphics[width=70mm]{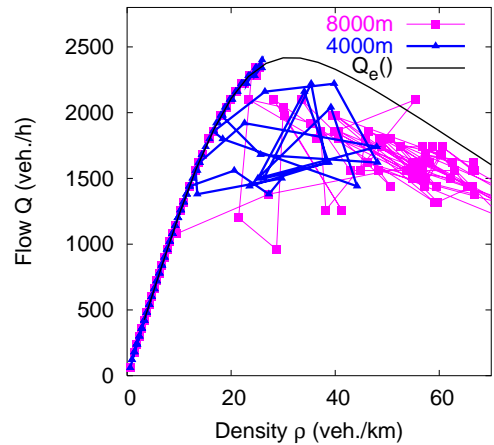} 
\end{center}
\caption[]{
Flow-density plot as in Fig.
\protect\ref{fig_fund}, simulated with the 
IDM without memory effects. To obtain the same degree of capacity and
stability, the parameters $T$ and $a$ have been changed to $1.05$ s and
$1$ m/s$^2$, respectively.
}
\label{fig_fundIDM}
\end{figure}
%
The question arises to which extent (i) the wide scattering,  
(ii) the distinct hysteresis effects indicated by the ratios
$Q\sub{max}/Q\sub{CT}$ and $Q\sub{max}/Q\sub{jam}$, and (iii) the low
values of the ``measured'' densities in congested and jammed traffic
are new features of the IDMM or occur 
 in the original IDM without memory effects as well.
Figure \ref{fig_fundIDM} shows a  simulation with the original IDM,
which is a special case of the IDMM for $\beta_T=1$.
The virtual detectors display scattering as well. However, the
hysteresis effects are much smaller and the density of congested
traffic is shifted to higher values, which are, in particular, higher than
the values usually observed in empirical data.

\section {Discussion}

\label{sec_diss}

We have modeled a memory effect in the behavior of drivers by coupling
existing car-following models to dynamical equations for some model parameters 
such as the minimum (safe) time gap, the desired velocity, or the typical acceleration. 
In this paper, we have used the IDM as underlying model
resulting in the IDMM, the ``IDM with memory''. 
The Gipps model \cite{Gipps81} seems to be  a suitable candidate as well.
It should be straightforward to apply the same concept to 
macroscopic models such as the GKT \cite{GKT,Treiber-Pinch} 
and to cellular automata.
In fact, the slow-to-start rule \cite{Barlovic} can be interpreted as
the special case of an instantaneous adaptation ($\tau=0$), 
which is only effective for standing traffic, i.e., 
the corresponding ``level-of-service function'', Eq. \refkl{lambda0},
would be given by $\lambda_0(v)=1$ for $v>0$ and $\lambda_0(v)=0$ for $v=0$. 

The concept could be also generalized to include the traffic density or the 
velocity variance in determining the level-of-service function.
This would allow to model different kinds of adaptation behavior to different types of congested
traffic such as homogeneous congested traffic (HCT) and oscillating congested
traffic (OCT) \cite{Phase}.

As is the case for the IDM, the
IDMM is a deterministic car-following model
with a unique fundamental diagram.
It has two new parameters, 
the adaptation factor $\beta_T$ and the adaptation time $\tau$, 
which can be estimated from single-vehicle data. 
Simulations with the IDMM
suggest that the adaptation of drivers to the surrounding
traffic happens on time scales of a few minutes and play, in fact, an important role in
explaining the inverse-$\lambda$ shape and the wide scattering of flow-density
data in the congested regime measured by stationary detectors.
Despite its simplicity, the model seems to be accurate enough to 
enable, for the first time, a direct analysis of the conventional
interpretation of macroscopic flow-density 
data with surprising results both for the
theoretician and the practitioner. 

For the theoretician, probably 
the most interesting result is the direct demonstration 
of a simple mechanism that explains 
the much-discussed wide scattering of congested traffic
by longitudinal traffic instabilities.
Important for the practitioner, the results suggest that 
the real traffic situation, in terms of the traffic density, 
is often worse than 
the macroscopic density data suggest. 
The resulting oscillations and even short periods of standing traffic are
hidden in the ``scattering'' of the data.
In contrast to previous results guided by 
theoretical considerations \cite{Ker02_PRE}
the simulations suggest that congested traffic is nearly always unstable.
This is supported by our analysis of more than 300 empirical examples of
congestion from various
freeways with a new visualization tool \cite{Treiber-smooth}. The majority of
all congested traffic patterns displayed stop-and-go waves, many of them with 
growing amplitudes while they travel upstream. Another factor obscuring 
the instability of real congested traffic
is the often observed {\it convective} stability, i.e., growing
perturbations can only propagate upstream, resulting in homogeneous congested
traffic of high density near the downstream front of congestion (``pinch effect'').

In our simulations, we have excluded most sources that would not surprise
to produce scattering: We have not assumed heterogeneous multi-lane traffic
exposed to fluctuation effects. Instead, we have assumed identical vehicles
on a single lane with a dynamics given by a deterministic model 
with a unique fundamental diagram.
We do not claim, however, that longitudinal instabilities
and the memory effect would be the only cause leading
to the observed scattering. Obviously, heterogeneous traffic plays a
role as well. 
Furthermore, the role of lane changes remains to be investigated.
Finally, it should be emphasized that we have investigated macroscopic
implications of a microscopic model. 
To explain microscopic statistical
properties such as the observed scaling law for the fluctuations of
sample-average time headways \cite{Katsu03-condmat}, one probably needs to
simulate both heterogeneous and multi-lane traffic.
Microscopic statistical properties will
be investigated in a forth-coming paper.

\medskip

{\bf Acknowledgments:}
The authors  would like to thank for financial support by the
DFG (grant No.~He 2789) and Arne Kesting for providing 
Fig. \ref{fig_Tdistr}.
We {are also grateful to} the 
{\em Dutch Ministry of Transport, Public Works
and Water Management} for providing the single-vehicle
induction-loop-detector data.


\end{document}